\documentclass[runningheads]{llncs}
\usepackage[T1]{fontenc}
\usepackage{graphicx}
\usepackage{soul}
\usepackage{color}
\usepackage{url}
\usepackage[hidelinks]{hyperref}
\usepackage[utf8]{inputenc}
\usepackage[small]{caption}
\usepackage{graphicx}
\usepackage{amsmath, amsfonts, amssymb}
\usepackage{multirow}
\usepackage{booktabs}
\usepackage[switch]{lineno}
\begin{document}
%
\title{How Homogenizing the Channel-wise Magnitude Can Enhance EEG Classification Model?}
\titlerunning{Channel-wise Magnitude Homogenization EEG Classification}
%
\author{Huyen Ngo\inst{1} \and
Khoi Do\inst{1} \and
Duong Nguyen\inst{2} \and
Viet Dung Nguyen\inst{1} \and
Lan Dang\inst{1}}

\authorrunning{H. Ngo et al.}

\institute{Hanoi University of Science and Technology, Pusan National University\\
\email{\{ngoanhhuyen, khoido, mduongbkhn, nvdung.bme, danghoanglan271\}@gmail.com}}
%
%
\maketitle              
\begin{abstract}
    A significant challenge in the electroencephalogram EEG lies in the fact that current data representations involve multiple electrode signals, resulting in data redundancy and dominant lead information. However extensive research conducted on EEG classification focuses on designing model architectures without tackling the underlying issues. Otherwise, there has been a notable gap in addressing data preprocessing for EEG, leading to considerable computational overhead in Deep Learning (DL) processes. In light of these issues, we propose a simple yet effective approach for EEG data pre-processing. Our method first transforms the EEG data into an encoded image by an Inverted Channel-wise Magnitude Homogenization (ICWMH) to mitigate inter-channel biases. Next, we apply the edge detection technique on the EEG-encoded image combined with skip connection to emphasize the most significant transitions in the data while preserving structural and invariant information. By doing so, we can improve the EEG learning process efficiently without using a huge DL network. Our experimental evaluations reveal that we can significantly improve (i.e., from $2\%$ to $5\%$) over current baselines.

\end{abstract}
\section{Introduction}
Electroencephalograms (EEGs) hold great potential for medical advancements, including identifying neurological disorders and enabling mind-computer interfaces \cite{Altaheri2021}, \cite{Demir2021}, \cite{Xu2023}. Nonetheless, decoding the complex patterns embedded in EEG data necessitates the utilization of artificial intelligence (AI). AI models can learn from and interpret vast datasets, unlocking secrets within brain waves. However, applying a deep learning model to EEG classification presents unique challenges due to the temporal and non-linear nature of EEG signals, which can lead to overfitting and unreliable results.

Researchers are working on refining model architecture and enhancing feature extraction to fully exploit the combined temporal and spatial nature of multi-channel EEG data. Recurrent neural networks (RNNs) \cite{Yang2020}, \cite{Najafi2022} and LSTMs \cite{Spampinato2017,Zheng2020,Zheng2021} and \cite{Fares2020} excel in capturing the temporal evolution of brain activity across channels, but they can overlook the crucial spatial aspect, which can lead to overfitting. Convolutional neural networks (CNNs) \cite{EinShoka2023}, \cite{Liu2023}, EEG-based architectures \cite{eegnet}, \cite{eegchannelnet} are better at extracting spatial features from multi-channel data, but their dependence on 1D convolution limits their ability to capture intricate temporal relationships between channels. Feature enhancement methods, such as siamese networks \cite{eegchannelnet}, \cite{Spampinato2017} learning a joint representation between EEG and visual stimuli, may not generalize well to other EEG-based classification problems. Converting EEG signals into grayscale heatmaps \cite{Ahmed2022} can also distort the temporal structure of the data, leading to information loss. Moreover, despite state-of-the-art models achieving progress in signal classification, the core challenge of multi-channel signals, particularly EEG, remains largely unaddressed \cite{Craik2019}, \cite{Ellis2023}. The intricate nature of EEG data, with amplitude and varying frequency across channels, can lead to overlapping signals and the subsequent masking of vital information in certain channels \cite{Khosla2020,JalalyBidgoly2020}.

The paper introduces a two-step approach to pre-process EEG signals, aiming to improve the accuracy of EEG signal classification. It uses Inverted Channel-wise Magnitude Homogenization (ICWMH) to transform 1D EEG signals into high-dimensional representations, while FEvSC enhances information density by extracting features beyond standard temporal dynamics. The method also introduces a skip connection mechanism, enriching the feature set and providing a more comprehensive picture of brain activity. Our contributions can be summarized as follows:
\begin{itemize}
    \item An inverted channel-wise magnitude homogenization is proposed to address the variance in signal amplitudes across channels, ensuring equalized contribution among channels. 
    \item A feature enrichment via skip connection leverages the sequential nature of EEG signals to extract additional features, especially long-range dependencies, enriching the overall feature set.
    \item Extensive experiments comparing our method to various baselines proved its effectiveness, and ablation tests were used to assess its adaptability.

\end{itemize}
\section{Methodology}

\subsection{Preliminary and Notations}
We consider a dataset comprising $N$ training samples $\{(\textbf{x}^i, \textbf{y}^i)\}_i^N$, with $\textbf{x}^i\in\mathbb{R}^{d_x}$ and $\textbf{y}^i\in\mathbb{R}^{d_y}$ denotes the $i^{th}$ EEG sampled signal and the corresponding ground truth. Each image $\textbf{x}^i$ comprises $H \times W$ pixels denoted by $\textbf{x}^i = \{x^i(a,b)\}$, with $a$ and $b$ as pixel indices.The dataset encompasses $M\sim p(M)$ classes, where $p(M)$ denotes the categories' probability distribution. For a sample belonging to class $m$ we have $(\textbf{x}_m^i, \textbf{y}_m^i)$, where $m\in\left[1, \dots, M\right]$. To achieve EEG classification, we employ a model parameterized by a weight matrix $\mathcal{W} = \left[w_1, \dots, w_M\right]$. The objective is to optimize this model by minimizing the Cross-Entropy Loss function (Equation \ref{eq:loss}), which effectively measures the discrepancy between the predicted class probabilities and the true class labels. The specific problem tackled in this work can be formally described by Equation (\ref{eq:seg-cls}). 
\begin{equation}
     \label{eq:seg-cls}
     \underset{\mathcal{W}\in \mathbb{R}^{d\times M}}{\min} \mathcal{L}(\mathcal{W}) = \frac{1}{N}\frac{1}{M}\sum_{i=1}^{N}\sum_{m=1}^M[\mathcal{F} (\textbf{x}_m^i,\textbf{y}_m^i\vert \mathcal{W})]
\end{equation}
Equation (\ref{eq:loss}) further specifies the Cross-Entropy loss function, which plays a central role in training models for classification tasks.
\begin{equation}
    \label{eq:loss}
     \mathcal{F} (\textbf{x}^i,\textbf{y}^i) = -\sum_{m=0}^{M-1}\log\left(\frac{
     \exp\{f_\mathcal{W}(\textbf{x}^i_m)\}}{\sum^{M-1}_{c=0}\exp\{f_\mathcal{W}(\textbf{x}^i_c)\}}\right)\textbf{y}^i_m
\end{equation}
where $f_\mathcal{W}: \mathbb{R}^{d_x} \rightarrow \mathbb{R}^{d_y}$ is a linear operator based weight model $\mathcal{W}$.
\subsection{Overall Architecture}
\begin{figure}
    \centering
    \includegraphics[width=1.0\linewidth]{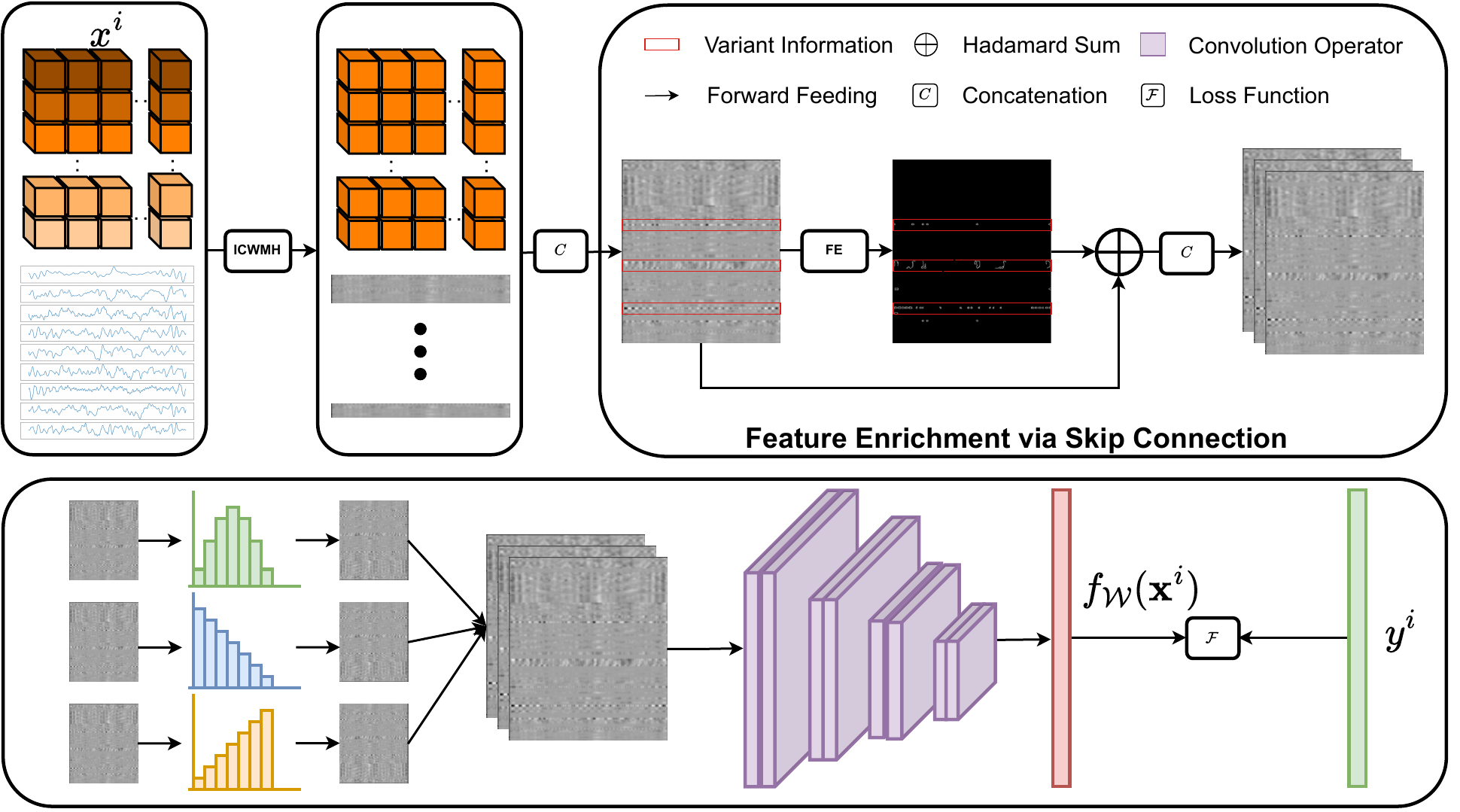}
    \caption{Overall two-step preprocessing methodology for EEG signal classification. \textbf{1) ICWMH: }an input EEG signal $x^i$ fed into the ICWMH process to normalize channel amplitudes and generate an encoded image with the size of  $ C \times L$ of the sample. \textbf{2) FEvSC: } FEvSC uses edge detection to extract useful variation information from an image. This extracted data is subsequently incorporated back into the encoded image. This potentially improves the performance of the EEG classification task.}
    \label{fig:diagram}
\end{figure}

The proposed methodology (Figure \ref{fig:diagram}) for EEG signal classification uses a two-step preprocessing approach to improve signal quality and feature extraction. This approach tackles the inherent variability in power distribution observed across EEG channels. First, the Inverted Channel-wise Magnitude Homogenization (ICWMH) technique is used to normalize signal amplitude, ensuring a balanced input for subsequent processing stages. This transformation encodes each EEG signal sample denoted as $x^i$ into an EEG-encoded image suitable for the CNN model. The encoded image is then enriched using a Feature Enrichment via Skip Connection (FEvSC) approach, which uses edge detection to capture variant information from the image. This information is then integrated back into the encoded image using the Hadamard sum, providing a more comprehensive representation of the CNN model. The enriched image, incorporating both original data and edge-derived features, is stacked into a three-layer structure, which is fed into the convolution operator of the CNN for classification, resulting in improved classification performance.

\subsection{Inverted Channel-wise Magnitude Homogenization (ICWMH)}
The input data $\textbf{x}^i \in \mathbb{R}^{C \times L}$ (refer to Figure \ref{fig:diagram}) is a composition of multiple stochastic random processes $x^i_c(t)$, where $C$, $L$ is the number of channels and channel signal length, respectively. The overall problem is maximizing the log-likelihood between the distribution of channel-wise signal $p(\textbf{y}^i|x^i_0, \dots, x^i_C)$ and the estimated class distribution $p(\textbf{x}^i | \textbf{y}^i)$ (refer to Equation \ref{eq:log}). Figure \ref{fig:diagram} illustrates the differences in power $\mathcal{P}$, or signal strength, across channels owing to the underlying neuronal activity and the amplitude of measured neural currents. Furthermore, the variance distribution of these signals ($p^i_k \neq p^i_h \ \forall \ h\neq k$) highlights how the dominant frequencies differ across brain regions. Consequently, the gradients across channels $\nabla_w\mathcal{F}(x_c^i)$ are dominated by the overwhelming one. 
\begin{equation}\label{eq:log}
    \mathcal{F}(\textbf{x}^i, \textbf{y}^i) = -\sum_{m=0}^{M-1}\log\left(p(\textbf{y}^i_m|x^i_0, \dots, x^i_C)\right)p(\textbf{x}^i_m | \textbf{y}^i_m)
\end{equation}

To address the issue of dominant magnitude across channels, ICWMH (see Figure \ref{fig:diagram}) aims to equalize signal magnitudes before training. This ensures each channel contributes equally, mitigating the dominance of specific channels by their overwhelming magnitude.
ICWMH employs a bottle-neck normalization process to squeeze the multi-channel signals within a fixed range. This technique serves a dual purpose which is removing redundant features within the local receptive field, preventing individual channels from dominating solely due to their magnitude, and balancing the contribution of each channel (represented by $\mathcal{P}(x^i_c)$ for channel $c$) by equalizing their influence on the learning process. Following this normalization, ICWMH leverages an interpolation step to enhance further the quality of information. This additional refinement ensures that while balancing the effect of overwhelming channels (indicated by a ratio of $\frac{\mathcal{P}(x^i_c)}{\sum_C\mathcal{P}(x^i_c)} \rightarrow 1$ ), crucial frequency properties are preserved (refer to Equation \ref{eq:lim}). 
\begin{equation}\label{eq:lim}
    \lim\|\nabla_\mathcal{W}\mathcal{F}(x^i_h) - \nabla_\mathcal{W}\mathcal{F}(x^i_k)\| \rightarrow 0, \forall h \neq k
\end{equation}
This translates to a learning process where no single channel holds undue sway, allowing for the extraction of richer and more informative features from the entire multi-channel spectrum. 

\subsection{Variant Feature Extractor}
\begin{figure*}[!ht]
    \centering
    \includegraphics[width = 0.9\textwidth]{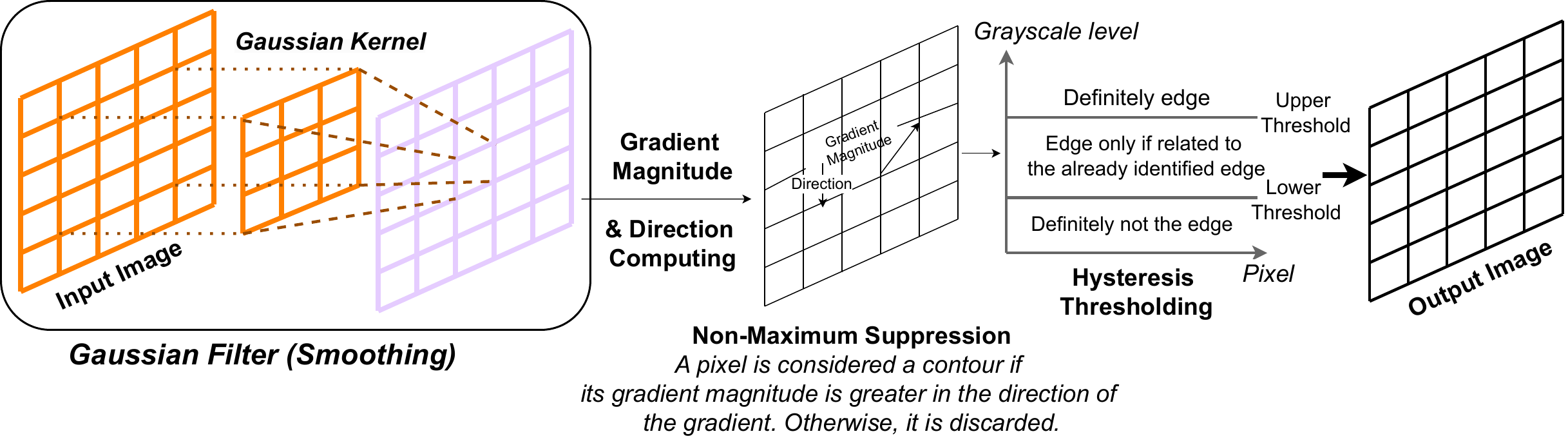}
    \caption{Overview of Variant Feature Extractor Method}
    \label{edge_detection}
\end{figure*}
In the EEG-encoded image, the salient features come from the change between image regions. Thus, by applying edge detection on the image (Figure \ref{edge_detection}), we can get the salient features with low computation complexity. To do so, we first define the pixel-wise gradient magnitude:
\begin{align}
    M(x^i(a,b)) = \sqrt{[\nabla_a x^i(a,b)]^2 + [\nabla_b x^i(a,b)]^2},
\end{align}
where $a$ and $b$ are pixel row and column indices, and the pixel-wise gradient direction is as follows: 
\begin{align}
    \theta (x^i(a,b)) = \arctan (\nabla_a x^i(a,b) / \nabla_b x^i(a,b))
\end{align}
Based on the two aforementioned features, we can find the salient representations as follows: 
\begin{align}
    \textrm{Edge}(x^i(a,b)) = \begin{cases}
    G(a, b), & \text{if } G(a, b) \geq G(a + \nabla a, b + \nabla b) \notag \\
             & \text{ and } G(a, b) \geq G(a - \nabla a, b - \nabla b), \\
    0, & \text{otherwise},
    \end{cases}
\end{align}
\subsection{Feature Enrichment via Skip Connection (FEvSC)}
The Feature Enrichment via Skip Connection (FEvSC) approach is proposed to improve the classification performance of an EEG model. It uses the strengths of the first proposed technique, ICWMH, and edge detection methods to enrich the feature space of EEG data. The approach as described in Equation \ref{eq:inv-var} aims to mitigate redundancy while preserving essential structural and variant features inherent in EEG signals, enhancing predictive accuracy.
\begin{equation}\label{eq:inv-var}
    \hat{x}^i = \textrm{ICWMH}(x^i) + \mathcal{I}(x^i)
\end{equation}
In this equation, $\mathcal{I}$ represents an affine transformation influenced by general edge detection principles. Edge detection in EEG signal analysis helps isolate significant boundaries and transitions, indicating critical neural activities and accentuating high-frequency components associated with subtle brain activities. This enhances contrast and clarity of signal features, highlighting areas of neural activity crucial for accurate classification. The FEvSC approach uses skip connections in its neural network architecture, allowing direct access to both raw and processed EEG signals. These enriched features provide refined input, aiding in learning complex patterns for accurate EEG classification. Combining edge detection and skip connections offers a promising direction for improving EEG classification system performance.

\section{Experiment}
\textbf{Datasets)} \textbf{Perceive Lab \cite{Spampinato2017}, \cite{eegchannelnet}. } The Perceive Lab dataset contains EEG responses from 6 subjects who viewed 2,000 unique objects (40 classes from ImageNet \cite{deng2009imagenet}) during 10 seconds to achieve 11,964 EEG segments. Each sample has 128 channels and 500 time-step data. The dataset is sampled in 50Hz 
 with notch filtering, channel-wise z-score normalized, and bandpass filtered across three frequency ranges. \textbf{High-gamma-dataset (HGD) \cite{HBM:HBM23730}.} The High-Gamma Dataset is a 128-electrodes dataset from 14 healthy subjects, consisting of 1000 four-second trials of executed movements divided into 13 runs per subject. The four classes of movements were left, right, feet, and rest. The datasets are each divided into training (80\%), validation (10\%), and test (10\%) sets.

\textbf{Models)} We reimplemented and evaluate LSTM networks, stacked bidirectional LSTMs \cite{Spampinato2017,Zheng2020,Zheng2021,Fares2020}, EEGNet \cite{eegnet} and EEGChannelNet \cite{eegchannelnet} model, Siamese networks \cite{eegchannelnet}, 2D EGG-encoded grayscale heatmaps\cite{Ahmed2022}. We trained these models under identical conditions with the learning rate of $9e-04$, the batch size is 64, and the optimizer is Adam under 100 epochs. This comprehensive exploration provided valuable insights into the effectiveness of different EEG classification approaches.

\textbf{Problems)} This study presents a two-step preprocessing method for enhancing feature representation. First, data is normalized using Inverted Channel-Wise Magnitude Homogenization (ICWMH). Second, the variant feature is extracted through edge detection, using Gaussian blur to reduce noise and employing Adaptive Edge and Canny Edge Detection techniques. 

\textbf{Hyperparamter Tuning)} In particular, the training configuration for Perceive Lab dataset employs Canny edge detection with thresholds (50, 120) in bilinear interpolation mode, along with a Gaussian blur kernel size of (3, 3).  For the HGD dataset, the baseline configuration utilizes adaptive edge detection with mean thresholding in bilinear interpolation mode and the same Gaussian blur kernel size.
\subsection{Comparison to State-of-the-art methods}
ICWMH with Edge Detection shown in Table \ref{table:eeg_model_accuracies}, achieves an accuracy of approximately 66\% on a dataset of 40-class images from the Perceive Lab challenge. This surpasses previous approaches, such as GIE (Grayscale Image Encoded), which achieved 64\% accuracy. The key improvement lies in its additional edge detection step, which extracts critical features within the EEG data, boosting accuracy to 65.78\%. This approach outperforms standard and specialized architectures like LSTMs and EEGNet by a large margin and surpasses EEChannelNet by nearly 20\%. ICWMH+ED's ability to extract richer information from EEG data paves the way for advancements in brain-computer interfaces and neurological disease diagnosis. When applied to the HGD dataset, ICWMH+ED achieves an accuracy of 57.18\%, double the accuracy of the grayscale-encoded image method without edge detection. This demonstrates the promise of ICWMH+ED in EEG classification tasks, even exceeding previous RNN models by an accuracy of 7\%.
\begin{table}[!ht]
\centering
\caption{Test accuracies of Different Models for EEG Classification on The Perceive Lab Dataset and The High Gamma Dataset: Baseline Parameters}
\resizebox{0.9\columnwidth}{!}{%
\begin{tabular}{@{}lcc@{}}
\toprule
\multicolumn{1}{c}{\multirow{3}{*}{\textbf{Method}}} & \multicolumn{2}{c}{\textbf{Dataset}} \\
\cmidrule{2-3}
\multicolumn{1}{c}{} & \textbf{Perceive Lab \cite{Spampinato2017} \cite{eegchannelnet}} & \textbf{HGD \cite{HBM:HBM23730}}  \\
\midrule
\multicolumn{1}{c}{} & \multicolumn{1}{c}{Accuracy(\%)} & \multicolumn{1}{c}{Accuracy(\%)} \\
LSTM \cite{Spampinato2017} & 23.97 $\pm$ 0.23 & 50.47 $\pm$ 0.14 \\
Stacked-BiLSTM \cite{Spampinato2017}  & 21.26 $\pm$ 0.26 & 50.58 $\pm$ 0.21 \\
EEGNet \cite{eegnet} & 30.00 $\pm$ 0.33 & - \\
EEGChannelNet \cite{eegchannelnet} & 39.58 $\pm$ 0.29 & -\\
Grayscale image encoded \cite{Ahmed2022} & 64.00 $\pm$ 0.24 & 24.54 $\pm$ 0.49 \\
\midrule
\textbf{ICWHM+ED (Ours)} & \textbf{65.78} $\pm$ \textbf{0.52} & \textbf{57.18} $\pm$ \textbf{0.47} \\
\bottomrule
\end{tabular}
}
\label{table:eeg_model_accuracies}
\end{table}
\subsection{Ablation Study}
\textbf{Interpolation Method)} The study examines the impact of different interpolation methods on EEG classification performance (refer to Table \ref{table:ablation test}), focusing on 'bilinear' and 'nearest' techniques. The 'bilinear' method \cite{Mastyo2013} achieves baseline accuracy for both datasets, while the 'nearest' method \cite{Rukundo2021} results in lower accuracy, especially in the HGD dataset, where the accuracy drops significantly to half of the baseline. The choice of interpolation significantly impacts the classification model's effectiveness. 

\textbf{Edge Threshold)} The study investigates different threshold values \cite{AkbariSekehravani2020}, \cite{Cao2020} in EEG data. The threshold of (50, 120) set Perceive Lab dataset achieves the highest accuracy (65.78\%). Lower threshold settings show a slight decrease in accuracy (64.74\%), while higher threshold settings result in a significant drop in accuracy (51.75\%), potentially omitting crucial information and leading to underfitting. The HGD dataset performs best with a threshold of (40, 120), achieving 56.81\% accuracy. A stricter threshold setting leads to lower accuracy, especially with an increased upper threshold, causing underfitting and a significant drop in accuracy. 

\textbf{Gaussian Blur Kernel)} The paper explores the impact of different Gaussian blur kernel sizes \cite{Liu2021} on EEG classification accuracy. The baseline kernel size of (3,3) achieves the highest accuracy at 65.78\% for Perceive Lab and 57.18\% for the HGD dataset. However, as kernel size increases to (5,5) and (7,7), accuracies decrease, possibly due to excessive smoothing, which can lead to a loss of critical signal detail. This highlights the importance of selecting an appropriate level of Gaussian blurring for EEG image preprocessing. 

\textbf{Edge Threshold)} The experiment also investigates the effectiveness of adaptive thresholding methods \cite{Liu2020} for edge detection. Adaptive Mean Thresholding achieves an accuracy of 62.98\% for Perceive Lab and baseline accuracy for the HGD dataset, effectively balancing local variations in illumination in EEG imaging. Adaptive Gaussian Thresholding, however, achieves lower accuracy for both datasets, suggesting a more localized approach to thresholding.

\begin{table}[!ht]
\centering
\caption{Performance of ablation tests on Perceive Lab and HGD datasets. 1) Interpolation method: 'bilinear' interpolation smooths transitions between pixel values, while 'nearest neighbor' interpolation preserves the original pixel values. 2) Canny edge threshold: Thresholds set at (40,120), (50, 100), (50, 120), and (50, 140) affect the quality of variant feature extraction. 3) Gaussian Blur Kernel: Tests on 3 kernel sizes (3,3), (5,5), and (7,7) for different levels of smoothing.  4) Adaptive Edge Threshold: Mean thresholding calculates the threshold for a pixel based on the average of a neighborhood area minus a constant, and Gaussian thresholding uses a Gaussian-weighted sum of neighborhood values.}
\resizebox{0.9\columnwidth}{!}{%
\begin{tabular}{lccc}
\toprule
\multicolumn{2}{c}{\textbf{Ablation Test}} & \multicolumn{2}{c}{\textbf{Dataset}} \\
\midrule
\multicolumn{1}{c}{\multirow{2}{*}{\textbf{Method}}} & \multicolumn{1}{c}{\multirow{2}{*}{\textbf{Parameters}}} & \textbf{Perceive Lab \cite{Spampinato2017}} & \textbf{HGD \cite{HBM:HBM23730}} \\
\cmidrule{3-4}
\multicolumn{1}{c}{} & \multicolumn{1}{c}{} & Accuracy(\%) & Accuracy(\%) \\
\midrule
\multirow{2}{*}{\textbf{Interpolation Method}} & 'bilinear' & \textbf{65.78 $\pm$ 0.52} & \textbf{57.18 $\pm$ 0.47} \\
 & 'nearest' & 55.86 $\pm$ 0.32 & 24.88 $\pm$ 0.41 \\
\midrule
\multirow{4}{*}{\textbf{Canny Edge Threshold}} & (40,120) & 48.13 $\pm$ 0.12 & 56.81 $\pm$ 0.38  \\
 & (50,100) & 64.74 $\pm$ 0.41 & 45.76 $\pm$ 0.38 \\
 & (50,120) & \textbf{65.78 $\pm$ 0.52} & 55.98 $\pm$ 0.23 \\
 & (50,140) & 51.75 $\pm$ 0.17 & 24.88 $\pm$ 0.16 \\
 \midrule
\multirow{3}{*}{\textbf{Gaussian Blur Kernel}} & (3,3) & \textbf{65.78 $\pm$ 0.52} & \textbf{57.18 $\pm$ 0.47} \\
 & (5,5) & 58.93 $\pm$ 0.31 & 55.45 $\pm$ 0.21 \\
 & (7,7) & 54.25 $\pm$ 0.27 & 52.53 $\pm$ 0.32 \\
 \midrule
\multirow{2}{*}{\textbf{Adaptive Edge Threshold}} & Mean Threshold & 62.98 $\pm$ 0.28 & \textbf{57.18 $\pm$ 0.47} \\
 & Gaussian Threshold & 60.49 $\pm$ 0.32 & 53.44 $\pm$ 0.41 \\
 \bottomrule
\end{tabular}%
}
\label{table:ablation test}
\end{table}
\section{Conclusion}
This paper presents a groundbreaking EEG classification method using Inverted Channel-wise Magnitude Homogenization (ICWMH) and Edge Detection. The method achieves approximately 66\% accuracy rate in 40 classes of classification tasks, highlighting the importance of improved feature representation and balanced channel input. The process converts EEG signals into expanded dimensional representations and integrates long-range dependencies, extracting a broader set of features. The study emphasizes the need for careful hyperparameter calibration and the delicate interplay between noise suppression and feature retention in EEG signal classification success. The methodology could serve as a new benchmark in the field.

\bibliographystyle{splncs04}
\bibliography{ref}

\clearpage
\appendix

\section{Related Works}
\subsubsection{Model Architecture Design}
Within the domain of EEG classification, significant research efforts have centered on optimizing established models like Recurrent Neural Networks (RNNs) and Convolutional Neural Networks (CNNs) for enhanced performance \cite{Iyer2022}. \cite{Spampinato2017} explored the potential of Long Short-Term Memory (LSTM) networks and their stacked variants, while \cite{Zheng2020} focused specifically on LSTMs-B with Swish activation and bagging ensembles. Further investigations by \cite{Fares2020} and \cite{Zheng2021} extended to BiLSTMs and attention-based models for visual object classification based on EEG signals. While RNNs excel at capturing temporal dynamics, concerns regarding potential overfitting due to limited spatial information extraction remain. This contrasts with CNNs, which demonstrate strong efficacy in EEG classification by effectively extracting relevant brain activity features. \cite{eegnet} solidified the value of CNNs in this domain through their compact EEGNet architecture specifically designed for EEG-based brain-computer interfaces. Further emphasizing the versatility of CNNs for EEG data, \cite{eegchannelnet} proposed EEGChannelNet, employing 1D CNNs for robust feature extraction. These advancements reflect the ongoing pursuit of improved performance and adaptability in EEG classification through continued model optimization and innovation.
\subsubsection{Feature Enhancement}
Prior research has significantly improved EEG feature data for better classification accuracy. \cite{eegchannelnet} and \cite{Spampinato2017}, \cite{Hajamohideen2023} pioneered the use of a Siamese network architecture to learn a joint embedding between EEG signals and images. This approach maximizes the similarity between embeddings from both modalities, thereby enhancing the model's representational power for EEG-based visual classification tasks. Further advancing EEG data utilization, \cite{Ahmed2022} introduced an innovative approach to transforming EEG signals into grayscale heatmap representations. This conversion from 1D signals to a 2D image format leverages the strengths of Convolutional Neural Networks (CNNs) by making relevant features more readily extractable for classification tasks. These advancements demonstrate the ongoing focus on enriching EEG feature data to unlock its full potential for accurate and reliable brain-computer interaction.

%

\end{document}